\documentclass{article}
\usepackage{spconf}

\usepackage{amsmath,amssymb,amsthm}
\usepackage{graphicx}
\usepackage{subfigure}
\usepackage{caption}
\usepackage{float}
\usepackage{bm}

\usepackage{cite}
\usepackage{booktabs,multirow}
\usepackage{makecell}
\usepackage{algorithm,algorithmicx,algpseudocode}

\usepackage{url}
\usepackage{verbatim}
\usepackage{tikz}
\usetikzlibrary{spy}

\newcommand{\x}		{\mathbf{x}}	
\newcommand{\y}		{\mathbf{y}}
\newcommand{\A}		{\mathbf{A}}
\newcommand{\z}	 {\mathbf{z}}
\newcommand{\W}	 {\mathbf{W}}
\renewcommand{\P}	 {\mathbf{P}}
\newcommand{\omg}	 {\mathbf{\Omega}}
\newcommand{\X}		{\mathbf{X}}	

\newcommand{\Z}		{\mathbf{Z}}
\newcommand{\0}	     {\mathbf{0}}
\newcommand{\R}	     {\mathbf{R}}
\newcommand{\rr}	     {\mathbf{r}}

\newcommand{\D}	 {\mathbf{D}}
\newcommand{\I}	     {\mathbf{I}}

\newcommand{\Rsf}	     {\mathsf{S}}

\newcommand{\diag}  {\mathsf{diag}}
\newcommand{\U}		{\mathbf{U}}	
\newcommand{\V}		{\mathbf{V}}
\newcommand{\Sig}		{\mathbf{\Sigma}}

\newcommand{\green}  {\textcolor{green}}

\title{Two-layer Residual Sparsifying Transform Learning for Image Reconstruction 	\vspace{-0.1in}}

\name{Xuehang Zheng$^1$, Saiprasad Ravishankar$^2$, Yong Long$^1$, Marc Louis Klasky$^3$, Brendt Wohlberg$^3$
\thanks{ This work was supported in part by the National Natural Science Foundation of China under Grant 61501292. \textsl{ (Corresponding author: Yong Long. Email: yong.long@sjtu.edu.cn))} }    
}
\address{$^1$University of Michigan - Shanghai Jiao Tong University Joint Institute, \\
	Shanghai Jiao Tong University, Shanghai 200240, China\\
$^2$Department of Computational Mathematics, Science and Engineering \\ and Department of Biomedical Engineering,
Michigan State University, East Lansing, MI 48824, USA\\
$^3$Los Alamos National Laboratory, Los Alamos, NM 87545, USA
\vspace{-0.15in}
}

\begin{document}
\ninept

\maketitle

\begin{abstract}
Signal models based on sparsity, low-rank and other properties have been exploited for image reconstruction from limited and corrupted data in medical imaging and other computational imaging applications. In particular, sparsifying transform models have shown promise in various applications, and offer numerous advantages such as efficiencies in sparse coding and learning. This work investigates pre-learning a two-layer extension of the transform model for image reconstruction, wherein the transform domain or filtering residuals of the image are further sparsified in the second layer. 
The proposed block coordinate descent optimization algorithms involve highly efficient updates. Preliminary numerical experiments demonstrate the usefulness of a two-layer model over the previous related schemes for CT image reconstruction from low-dose measurements.
\end{abstract}

\vspace{-0.05in}
\begin{keywords}
Low-dose CT, statistical image reconstruction, sparse representation, transform learning, unsupervised learning.
\end{keywords}


\vspace{-0.1in}
\section{Introduction}
\vspace{-0.05in}
\label{sec:intro}

Methods for image reconstruction from limited or corrupted data often exploit various inherent properties or models of the images. A variety of models such as sparsity, tensor, manifold, and convolutional models, etc. \cite{aharon:06:ksa, unitemri, zhang:17:tbd, garciacardona:18:cdl}, have been exploited for explaining or reconstructing images in computational imaging applications.
In this work, we focus our investigations on generalization of a subset of models called sparsifying transform models \cite{ravishankar:13:lst,ravishankar:15:lst}, their learning, and application to low-dose computed tomography (CT) image reconstruction.

A major challenge in CT imaging is to reduce the radiation exposure to patients while maintaining the high quality of reconstructed images. This is typically done by reducing the X-ray dose to low or ultra-low levels or by reducing the number of projection views (sparse-view CT). In such cases, conventional filtered back-projection (FBP) \cite{feldkamp:84:pcb} reconstructions suffer from artifacts that degrade image quality.

Model-based image reconstruction methods produce accurate reconstructions from reduced dose CT measurements  \cite{fessler:00:sir}. In particular, penalized weighted-least squares (PWLS) approaches, which have shown promise for CT reconstruction optimize a weighted-least squares data fidelity or measurement modeling term (for the logarithm of the measurements) along with added regularization exploiting prior knowledge of the underlying object \cite{thibault:06:arf}.

Learning signal models or priors from datasets of images or image patches is an attractive way to obtain adaptive CT image features to improve reconstruction. Recent works have proposed learning various models, including dictionary and sparsifying transform models \cite{xu:12:ldx, zheng:16:ldc, zheng:18:pua}, as well as supervised learning for reconstruction \cite{jin:17:dcn}. 
The learning of sparsifying transform models offers numerous advantages \cite{ravishankar:13:lst} over synthesis dictionary models. 
In particular, sparse coding in the dictionary model can be expensive, whereas in the sparsifying transform (ST) model, sparse coefficient maps are computed \emph{exactly and cheaply} by thresholding-type operations (i.e., transform sparse coding even with the $\ell_0$ “norm” is not NP-hard). Thus, transform learning-based approaches, including those for image reconstruction, can offer significant computational benefits \cite{zheng:18:pua} and often come with convergence guarantees \cite{ravishankar:15:lst, sabsam, ye:19:sld}.
Recent work has also shown that they can generalize better to unseen data than supervised deep learning schemes \cite{zheng:17:l1st, ye:19:sld}.

In this work, we investigate the model-based learning of a two-layer extension of the transform model \cite{ravishankar:18:lml} from datasets for image reconstruction. 
The transform domain or filtering residuals for the data are further sparsified in the second layer. 
The method in \cite{ravishankar:18:lml} exploited downsampling/pooling operations for image denoising that cannot be readily incorporated in the general inverse problem optimization explored here. 
Here, we propose pre-learning the two-layer transform (and estimating corresponding sparse coefficient maps) in a model-based fashion to minimize the aggregated transform domain residuals in the second layer, which is used as a regularizer for reconstruction. 
An efficient block coordinate descent algorithm is derived for learning and for reconstruction with the pre-learned regularizer. 
Unlike the recent multi-layer convolutional sparse coding (ML-CSC) approach~\cite{zeiler:10:dn, papyan:17:cnn, sulam:18:mcs}, which uses the general synthesis dictionary model and involves expensive sparse coding, exact and cheap sparse coefficients can be computed in our models. Moreover, ML-CSC sparsified the sparse coefficients over layers rather than reducing the modeling residuals. With the transform model, optimizing the residuals significantly improved performance. Finally, ML-CSC has not been investigated for imaging inverse problems.
Here, we present numerical experiments demonstrating potential for our approach for low-dose CT reconstruction compared to recent learned single layer transform and nonadaptive methods.

\vspace{-0.1in}
\section{Learning and Reconstruction Formulations}
\label{sec:formulation}
\vspace{-0.1in}

This section discusses the proposed two-layer framework and formulations for learning and image reconstruction.

\vspace{-0.15in}
\subsection{Two-Layer Residual Transform Learning}

For a signal $x \in \mathbb{R}^{N_p}$ and operator $W \in \mathbb{R}^{p \times N_p}$, the sparsifying transform model suggests that $Wx \approx z$, where $z$ has many zeros. Given the signal $x$ and operator $W$, the \emph{transform sparse coding} problem finds the best sparse approximation $z$ by minimizing the approximation error or residual in $Wx \approx z$, and the solution is obtained in closed-form by thresholding $Wx$ \cite{ravishankar:13:lst}.
When the transform is applied to all the overlapping patches of the image, the model is equivalent to a sparsifying filterbank for images \cite{ravishankar:18:lml, pfisbres:19:lfb}.

Here, we study a two-layer extension of the transform model, in which the transform domain residuals or sparse approximation errors in the first layer are further sparsified in the second layer.
We propose a patch-based formulation for learning, which could also be equivalently cast in a convolutional form \cite{ravishankar:18:lml}.
Given $N'$ vectorized (2D or 3D) image patches extracted from a dataset of CT images or volumes, we learn transforms $\{\omg_1, \omg_2 \in \mathbb{R}^{p \times p} \}$ by solving the following training optimization problem:
\vspace{-0.1in}
	\begin{equation}\label{eq:P0}
		\begin{aligned}
			& \min_{\{\omg_l,\Z_l\}}  \sum_{l=1}^{2}  \bigg \{\| \omg_l \R_l - \Z_l\|^2_F    +  \eta_l^2    \|\Z_l\|_0  \bigg \}   &\\
			& \;\;\; \mathrm{s.t.} \; \R_2 =   \omg_{1} \R_1 - \Z_1,  \,  \omg_l^T \omg_l   = \I \, \forall \,  l \, ,
		\end{aligned}
		\tag{P0}
		\vspace{-0.05in}
	\end{equation}	
	where $\R_1 \in \mathbb{R}^{p \times N'}$ denotes the matrix whose columns are the initial vectorized training image patches, $\R_2 \in \mathbb{R}^{p \times N'}$ denotes the residual maps in the second layer, and $\{\Z_1, \Z_2 \in \mathbb{R}^{p \times N'}\}$ denote sparse coefficient maps in two layers.
The non-negative parameters $\{\eta_l \} $ control the sparsity of the coefficient maps, with the $\ell_{0}$ ``norm'' counting the number of non-zero entries in a matrix or vector. The transforms $\{\omg_1, \omg_2\}$ are assumed to be unitary \cite{ravishankar:15:lst}, which simplifies the optimization, and $\I$ denotes the identity matrix.


\vspace{-0.15in}
\subsection{CT Image Reconstruction Formulation}

We propose using a pre-learned two-layer transform model as a prior for image reconstruction.
We reconstruct the image or volume $\x \in \mathbb{R}^{N_p}$ from noisy sinogram data $\y \in \mathbb{R}^{N_d}$  by solving the following PWLS optimization problem: 
\vspace{-0.1in}
\begin{equation}\label{eq:P1}	
	\min_{\x \geq \0}  \frac{1}{2}\|\y - \A \x\|^2_{\W}  + \beta \Rsf(\x),   
	\tag{P1}
	\vspace{-0.05in}
\end{equation}
where $\A \in  \mathbb{R}^{  N_d  \times N_p}$ is the system matrix of the CT scan, $\W = \diag \{w_i\} \in \mathbb{R}^{N_d \times N_d}$ is the diagonal weighting matrix with elements being the estimated inverse variance of $y_i$ \cite{thibault:06:arf}, parameter $\beta>0$ controls the trade-off between noise and resolution, and the regularizer $\Rsf (\x)$ based on (P0) is
\vspace{-0.1in}
\begin{equation}\label{eq:Rx_DST}
	\begin{aligned}
		\Rsf(\x) \triangleq  \min_{\{\Z_l\}}  \sum_{l=1}^{2}  \bigg\{  \|\omg_l \R_l - \Z_l\|^2_F + \gamma_l^2\|\Z_l\|_0   \bigg\}  & \\
		\mathrm{s.t.} \;  \R_2 =   \omg_1 \R_1 - \Z_1, \, \rr_1^j = \P^j\x  \, \forall \, j \, .
	\end{aligned}
		\vspace{-0.1in}
\end{equation}	
Here, $\{\gamma_l \} $ are non-negative scalar parameters, the operator $\P^j \in \mathbb{R}^{p\times N_p}$ extracts the $j$th patch of $p$ voxels of $\x$ as $\P^j \x$, and $\rr^j_1$ denotes the $j$th column of $\R_1$. The columns of $\Z_l $ are $\{\z_l^j \in \mathbb{R}^{p}\}_{j=1}^{N_{r}}$  and denote the transform-sparse coefficients in the $l$th layer, where $N_r$ is the number of extracted patches.

\vspace{-0.1in}
	\section{Algorithms}
\label{sec:algorithm}



\vspace{-0.1in}   
\subsection{Algorithm for Learning}
\label{subsec:learning_alg}
\vspace{-0.05in}   
We solve \eqref{eq:P0} using an \emph{exact} block coordinate descent algorithm that alternates between \textit{sparse coding steps} (solving for $\Z_1$ or $\Z_2$) and \textit{transform update steps} (solving for $\omg_1$ or $\omg_2$). 
The transforms $\omg_1$ and $\omg_2$ and the coefficients $\Z_2$ need to be first initialized. 
In our experiments, we used the 2D DCT and identity matrices to initialize $\omg_1$ and $\omg_2$ respectively, and the initial $\Z_2$ was an all-zero matrix.

\vspace{-0.15in}
\subsubsection{Sparse Coding Step for $\Z_1$}
Here, we solve the following sub-problem for $\Z_1$ with all other variables fixed: 
\vspace{-0.1in}   
\begin{equation}\label{eq:z1}
	\min_{\Z_1}    \|\omg_1 \R_1  - \Z_1\|^2_F    +    \| \omg_2 \R_2 - \Z_2\|^2_F   +   \eta_1^2   \|\Z_1\|_0 \, .
	\vspace{-0.05in}   
\end{equation}		
Substituting $\R_2 = \omg_1 \R_1 - \Z_1$ and using the unitary property of $\omg_2$, we rewrite \eqref{eq:z1} as  $ \min_{\Z_1}   2 \|\Z_1 - (\omg_1 \R_1 - 0.5 \omg_2^T\Z_2) \|^2_F  +  \eta_1^2   \|\Z_1\|_0 $. 
Then the optimal solution is obtained as $ \hat{\Z}_1 =  H_{\eta_1/\sqrt{2}} (\omg_1 \R_1 - 0.5 \omg_2^T\Z_2) $, where the \textit{hard-thresholding} operator $H_{\eta}(\cdot)$ zeros out vector entries with magnitude less than $\eta$.

\vspace{-0.15in}
\subsubsection{Transform Update Step for $\omg_1$}
With $\omg_2$, $\Z_2$, and $\Z_1$ fixed, we update $\omg_1$ by solving the following sub-problem:
\vspace{-0.05in}
\begin{equation}\label{eq:omega1}
		\min_{\omg_1}    \|\omg_1 \R_1  - \Z_1\|^2_F   +    \| \omg_2 (\omg_1 \R_1 - \Z_1) - \Z_2\|^2_F 
		\;\; \mathrm{s.t.} \; \omg_1^T \omg_1  = \I \, .
			\vspace{-0.05in}   
\end{equation}	
This is equivalent to minimizing the cost $2 || \omg_1 \R_1 - (\Z_1 + 0.5 \omg_2^T \Z_2)||_F^2$. 
Denoting the full singular value decomposition (SVD) of $ \R_1\Z_1^T + 0.5\R_1 \Z_2^T \omg_2$ as $\U_1 \Sig_1 \V_1^{T}$ (cf.~\cite{ravishankar:15:lst}), the optimal solution to \eqref{eq:omega1} is $ \hat{\omg}_1 = \V_1\U_1^T$.

\vspace{-0.15in}
\subsubsection{Sparse Coding Step for $\Z_2$}
With $\omg_2$, $\Z_1$, and $\omg_1$ fixed, we update $\Z_2$ by solving the following sub-problem:
\vspace{-0.05in}   
\begin{equation}\label{eq:z2}
	\begin{aligned}
		\min_{\Z_2}     \| \omg_2 (\omg_1 \R_1 - \Z_1) - \Z_2\|^2_F   +   \eta_2^2   \|\Z_{2}\|_0 \, .
	\end{aligned}
	\vspace{-0.05in}   
\end{equation}	
The optimal sparse coefficients for the second layer are readily computed in closed-form by hard-thresholding as $\hat{\Z}_{2} =  H_{\eta_2} (\omg_2 (\omg_1 \R_1 - \Z_1)  ) $.

\vspace{-0.15in}
\subsubsection{Transform Update Step for $\omg_2$}
Here, we update $\omg_2$ keeping the other variables fixed by solving:
\vspace{-0.05in}   
\begin{equation}\label{eq:omega2}
	\min_{\omg_2}    \| \omg_2 (\omg_1 \R_1 - \Z_1) - \Z_2\|^2_F
	\quad \mathrm{s.t.}  \quad \omg_2^T \omg_2  = \I \, . 
		\vspace{-0.05in}   
\end{equation}	
Denoting the full SVD of $  (\omg_1\R_1 - \Z_1) \Z_2^T$ as $\U_2 \Sig_2 \V_2^{T}$, the optimal solution to \eqref{eq:omega2} is $ \hat{\omg}_2 = \V_2\U_2^T$.

\vspace{-0.15in}	
\subsection{Image Reconstruction Algorithm}

We propose an alternating-type algorithm for \eqref{eq:P1} that alternates between updating $\x$ (\textit{image update step}), and $ \Z_1$ and $\Z_2$ (\textit{sparse coding steps}).  	

 \vspace{-0.15in}
\subsubsection{ Image Update Step}

With the variables $\Z_1$ and $\Z_2$ fixed, we solve \eqref{eq:P1} for $\x$, which reduces to the following weighted least squares problem:
 \vspace{-0.05in}   
\begin{equation}
	\label{eq:image}
	\min_{\x \geq \0} \frac{1}{2} \|\y - \A\x \|^2_{\W} + \Rsf_2(\x), \,
	\vspace{-0.05in}   
\end{equation}
where $\Rsf_2(\x) \triangleq \beta  \sum_{j=1}^{N_r} \big\{   \|\omg_2 (\omg_1\P^j \x - \z_1^j) - \z_{2}^j\|^2_2  + \|\omg_1 \P^j \x - \z_1^j\|^2_2 \big\}$.
We solve \eqref{eq:image} using the \emph{efficient} relaxed OS-LALM algorithm \cite{nien:16:rla}. 
The algorithmic details are similar to those in \cite{zheng:18:pua}.
	We precompute a diagonal majorizing matrix of the Hessian of the regularizer $\Rsf_2(\x)$ as $\D_{\Rsf_2}  \triangleq  \nabla^2 \Rsf_2(\x) = 4 \beta \sum_{j=1}^{N_r}  (\P^j)^{T}\P^j$.

\vspace{-0.1in}
\subsubsection{Sparse Coding Steps}
First, with $\x$ and $\Z_2$ fixed and $\X$ denoting the matrix with $\P^j \x$ as its columns, we update $\Z_1$ by solving
 \vspace{-0.05in}   
\begin{equation}\label{eq:recon_z1}		
	\min_{\Z_{1}}  \|\omg_1 \X - \Z_{1}\|^2_F + \gamma_1^2\|\Z_{1}\|_0 + \|\omg_2 (\omg_1\X - \Z_{1}) - \Z_{2}\|^2_F \, .
	 \vspace{-0.05in}   
\end{equation}	
Similar to the solution for \eqref{eq:z1}, the optimal solution for \eqref{eq:recon_z1} is $ \hat{\Z}_{1} =  H_{\gamma_1/\sqrt{2}} (\omg_1 \X - 0.5 \omg_2^T \Z_2  )$.

Next, with $\X$ and $\Z_1$ fixed, coefficients $\Z_2$ are updated by solving the following sub-problem:
\vspace{-0.05in}
\begin{equation}\label{eq:recon_z2}		
	\min_{\Z_{2}}  
	\|\omg_2 (\omg_1 \X - \Z_{1}) - \Z_{2}\|^2_F  +  \gamma_2^2\|\Z_{2}\|_0
	\, .
	\vspace{-0.05in}   
\end{equation}	
The optimal solution is $\hat{\Z}_{2} =  H_{\gamma_2} (\omg_2 (\omg_1 \X - \Z_{1})  )$.

\vspace{-0.1in}
\section{Experimental Results}
\label{sec:result}

We evaluated the proposed PWLS reconstruction method with a two-layer learned regularizer (referred to as \textbf{PWLS-MRST2}) and compared its image reconstruction quality with those of the \textbf{FBP} method with a Hanning window, and the \textbf{PWLS-EP} method that uses a non-adaptive edge-preserving regularizer $\Rsf(\x) = \sum_{j  =1}^{N_p} \sum_{k\in N_{j}}\kappa_{j} \kappa_{k} \varphi(x_j - x_k)$, where $N_j$ is the size of the neighborhood, $\kappa_j$ and $\kappa_k$ are the parameters encouraging uniform noise \cite{cho:15:rdf}, and ${\varphi}{(t)}\triangleq\delta^2 (  | t/\delta | - \log(1+| t/\delta |) )$ with $\delta=10$ Hounsfield units (HU)\footnote{Modified HU is used, where air is $0$ HU and water is $1000$ HU.}. We optimized the PWLS-EP problem using the relaxed OS-LALM method \cite{nien:16:rla}. We also compared to the previous \textbf{PWLS-ST} method that uses a learned single-layer (square) transform \cite{zheng:18:pua, zheng:16:ldc}. 

Various methods are compared quantitatively using the Root Mean Square Error (RMSE) and Peak Signal to Noise Ratio (PSNR) metrics in a region of interest (ROI). 
The RMSE of the reconstruction $\hat{\x}$ is defined as \mbox{RMSE $= \sqrt{\sum_{i=1}^{N_{p}}(\hat{x}_i-x^*_i)^2/{N_{p}}}$}, where $\x^*$ is the ground truth image and $N_{p}$ is the number of pixels (voxels) in the ROI.
We tuned the parameters of various methods for each experiment to achieve the lowest RMSE and highest PSNR.

\begin{figure}[!t]
	\centering  	
	\begin{tabular}{cc}
		\includegraphics[width=0.16\textwidth]{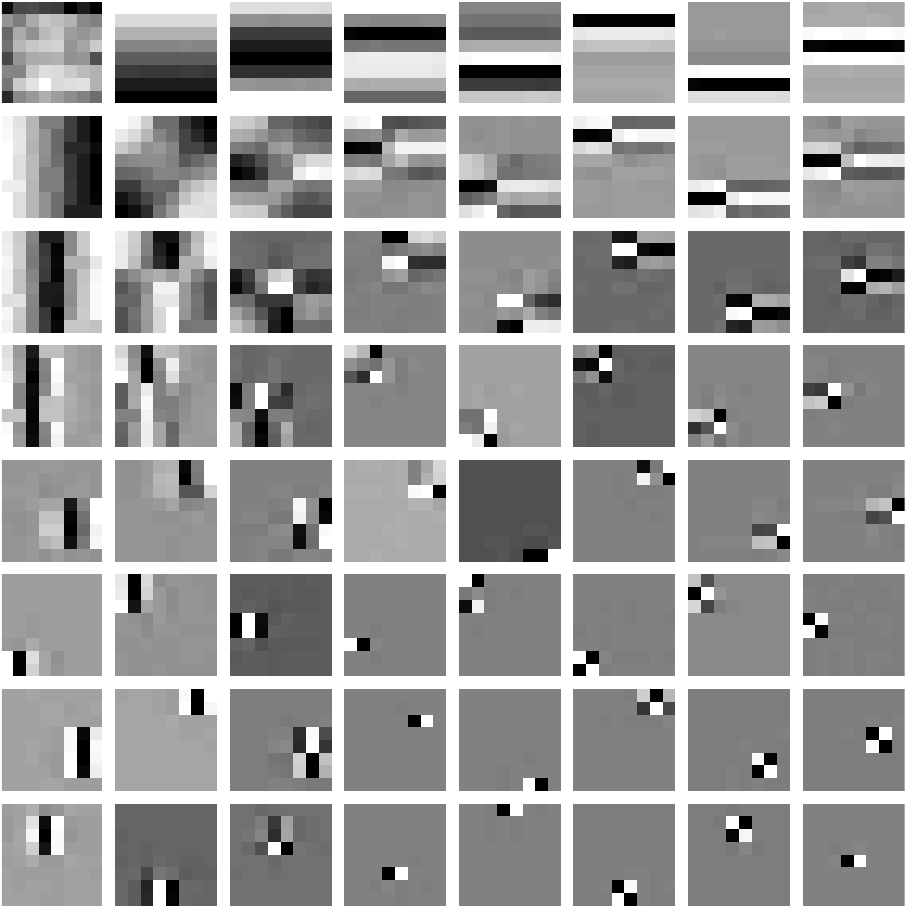}	&
		\includegraphics[width=0.16\textwidth]{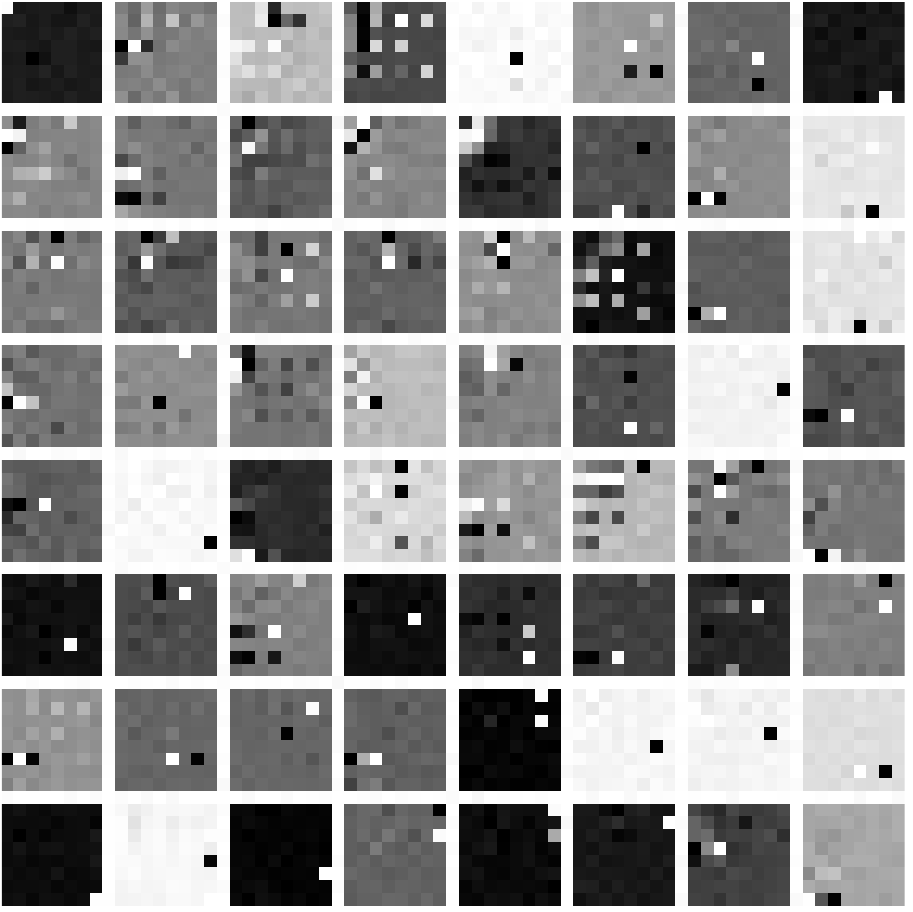}	\\
		(a) & (b)
	\end{tabular}
	\vspace{-0.1in}
	\caption{ Pre-learned sparsifying transforms $\omg_1$ (a) and $\omg_2$ (b) with $\eta_1 =  80$ and $\eta_2 =  60$. The rows of $\omg_1$ and the rows (1D atoms) of $\omg_2$ are reshaped into $8 \times 8$ patches for display.} 
	\label{fig:lear_tran}
	\vspace{-0.05in}
\end{figure}

We pre-learned two transforms (Fig.~\ref{fig:lear_tran}) for the proposed two-layer model from $8 \times 8$ image patches extracted from five $512 \times 512$ XCAT phantom \cite{segars:08:rcs} slices, with $\eta_1 = 80$, $\eta_2 = 60$, and a patch extraction stride $1 \times 1$. We ran $1000$ iterations of the learning algorithm in Section \ref{subsec:learning_alg} to ensure convergence.
We simulated 2D fan-beam CT test scans using $840 \times 840$ XCAT phantom slices (air cropped) that differ from the training slices, with pixel size $\Delta_x=\Delta_y=0.4883$ mm. Noisy sinograms of size $888 \times 984 $ were numerically simulated with GE LightSpeed fan-beam geometry corresponding to a monoenergetic source with $10000$, $5000$, and $3000$ incident photons per ray and no scatter, respectively. We reconstructed two $420 \times 420$ images with a coarser grid, where $\Delta_x=\Delta_y=0.9766$ mm. 
The ROI here was a circular (around center) region containing all the phantom tissues.

Initialized with FBP reconstructions, we ran the PWLS-EP algorithm for $50$ iterations using relaxed OS-LALM with $24$ subsets. 
The PWLS-EP result was used to initialize the adaptive methods. The parameters for different methods for $I_0=10000$, $5000$, and $3000$ are as follows: $\beta = 2^{16}, 2^{16.5}$, and $2^{16.5}$ respectively, for Slice~$1$ and $\beta=2^{16}$ for Slice~$2$ for PWLS-EP; $ (\beta, \gamma_1) =  \left ( 2 \times 10^5, 20 \right )$ , $\left ( 1.3 \times 10^5, 20 \right )$, and $\left ( 1.3 \times 10^5, 20 \right )$ respectively for Slice~$1$ and  $\left ( 2.2 \times 10^5, 20 \right )$, $\left ( 2 \times 10^5, 20 \right )$, and $\left ( 1.5 \times 10^5, 20 \right )$ for Slice~$2$ for PWLS-ST; and $(\beta, \gamma_1, \gamma_2) = \left ( 9 \times 10^4, 30, 10  \right )$, $\left (   4 \times 10^4, 30, 12 \right )$ and $\left (  3.5 \times 10^4, 30, 12 \right )$, respectively for Slice~$1$ and  $\left ( 8 \times 10^4, 30, 12  \right )$, $\left (   5 \times 10^4, 30, 12 \right )$ and \mbox{$\left (  5 \times 10^4, 30, 7 \right )$}  for Slice~$2$ for PWLS-MRST2.
For PWLS-ST and PWLS-MRST2, the image reconstruction algorithms were run for $1000$ and $1500$ outer iterations with $4$ and $2$ ordered subsets, respectively, and $2$ inner iterations of the image update step that ensured convergence.

Table~\ref{tab:2d} summarizes the RMSE and PSNR values for reconstructions with FBP, PWLS-EP, PWLS-ST, and the proposed PWLS-MRST2 for the three tested photon intensities. The adaptive PWLS methods significantly outperform the conventional FBP and the non-adaptive PWLS-EP. Moreover, PWLS-MRST2 with a learned two-layer model improves the reconstruction quality over the single-layer PWLS-ST scheme. It differs from PWLS-ST by only an additional simple sparse coding step and thus has a similar computational cost.

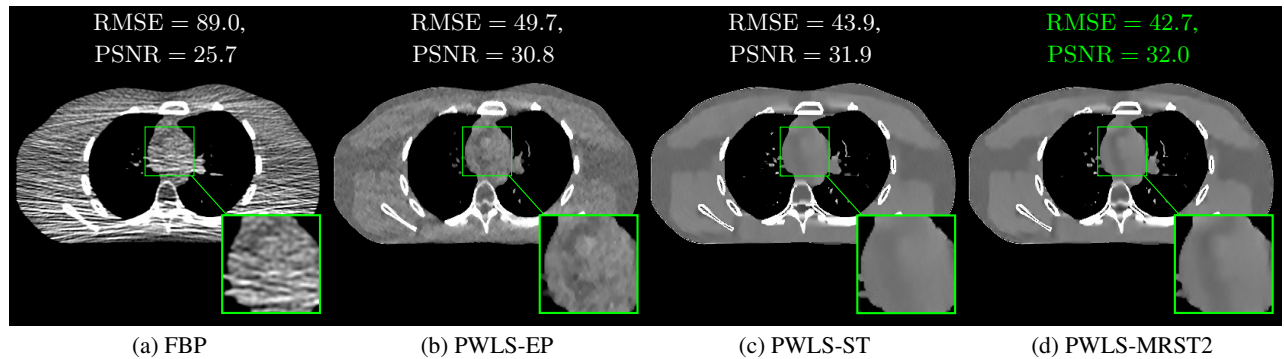
\begin{figure*}[!t]
	\centering
	\begin{tabular}{cccc}
		\begin{tikzpicture}
		[spy using outlines={rectangle,green,magnification=2,size=13mm, connect spies}]
		\node {	\includegraphics[width=0.24\textwidth]{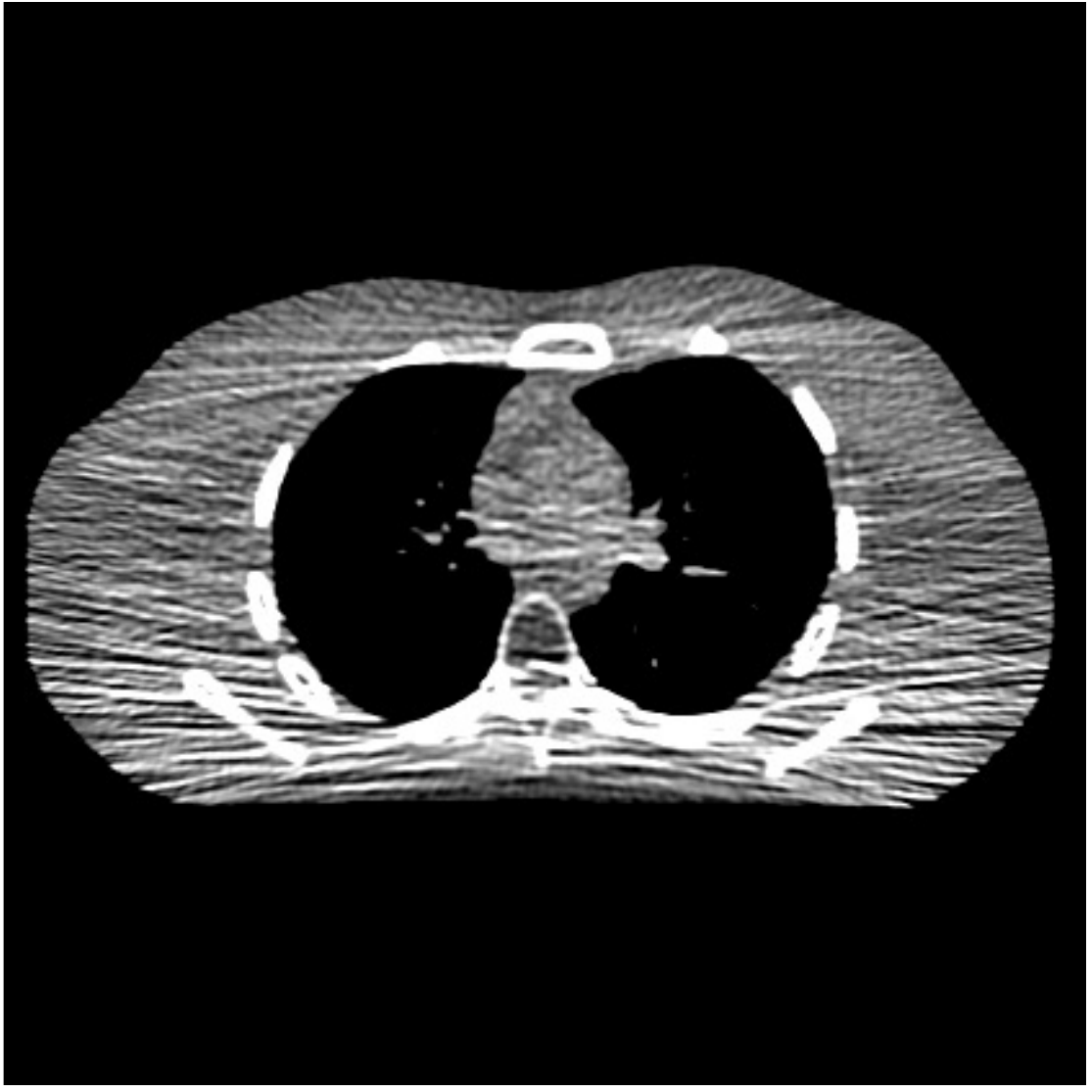}		};
		\spy on (0,0.2) in node [left] at (2,-1.3);	
		\node [white] at (0,1.9) {\small $\mathrm{RMSE} = 89.0$,};
		\node [white] at (-.05,1.5) {\small $\mathrm{PSNR} = 25.7$};

		\end{tikzpicture} 	&  \hspace{-0.3in}  \vspace{-0.05in}
		\begin{tikzpicture}
		[spy using outlines={rectangle,green,magnification=2,size=13mm, connect spies}]
		\node {	\includegraphics[width=0.24\textwidth]{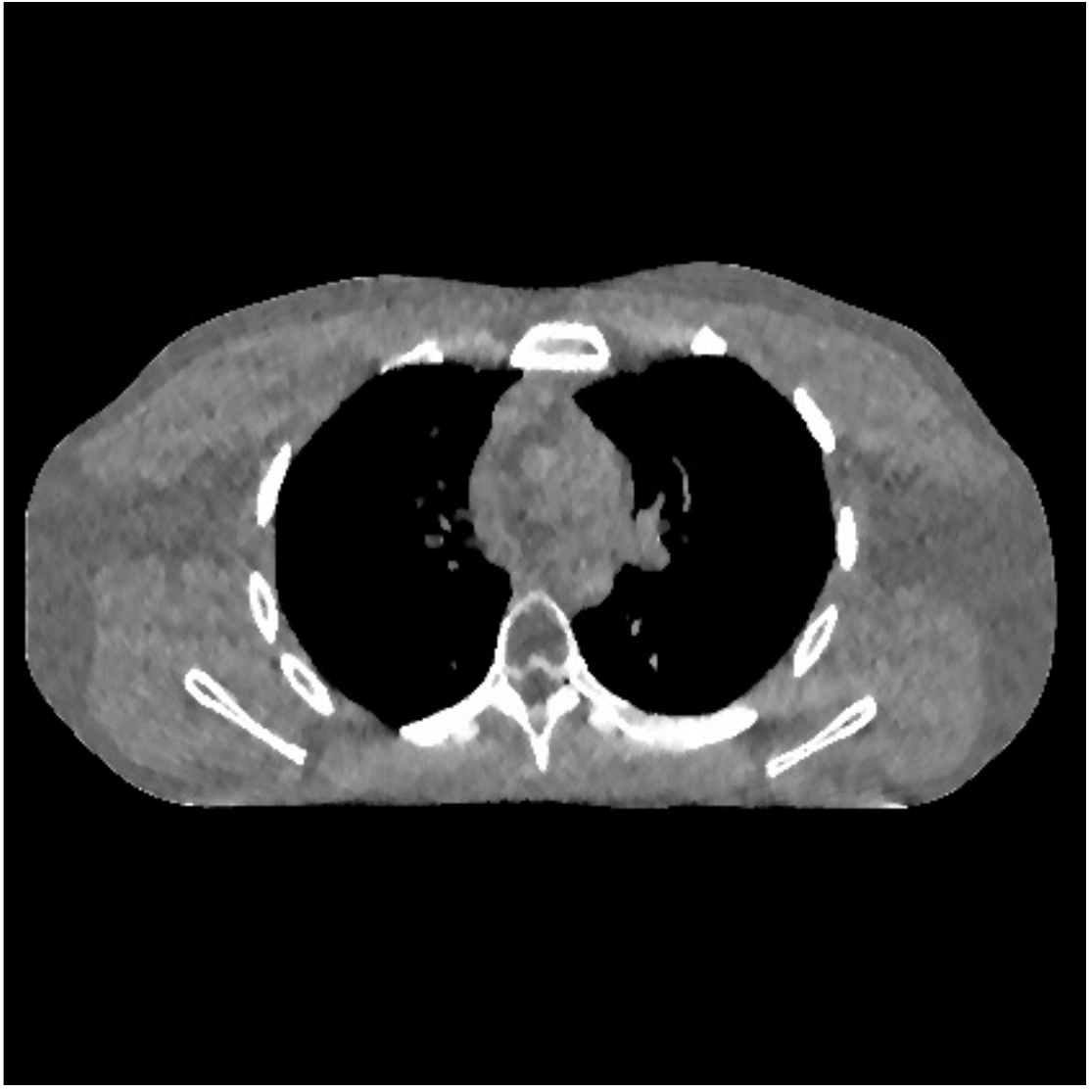}		};
		\spy on (0,0.2) in node [left] at (2,-1.3);	
		\node [white] at (0,1.9) {\small $\mathrm{RMSE} = 49.7$,};
		\node [white] at (-.05,1.5) {\small $\mathrm{PSNR} = 30.8$};
		\end{tikzpicture} &  \hspace{-0.3in}
		\begin{tikzpicture}
		[spy using outlines={rectangle,green,magnification=2,size=13mm, connect spies}]
		\node {	\includegraphics[width=0.24\textwidth]{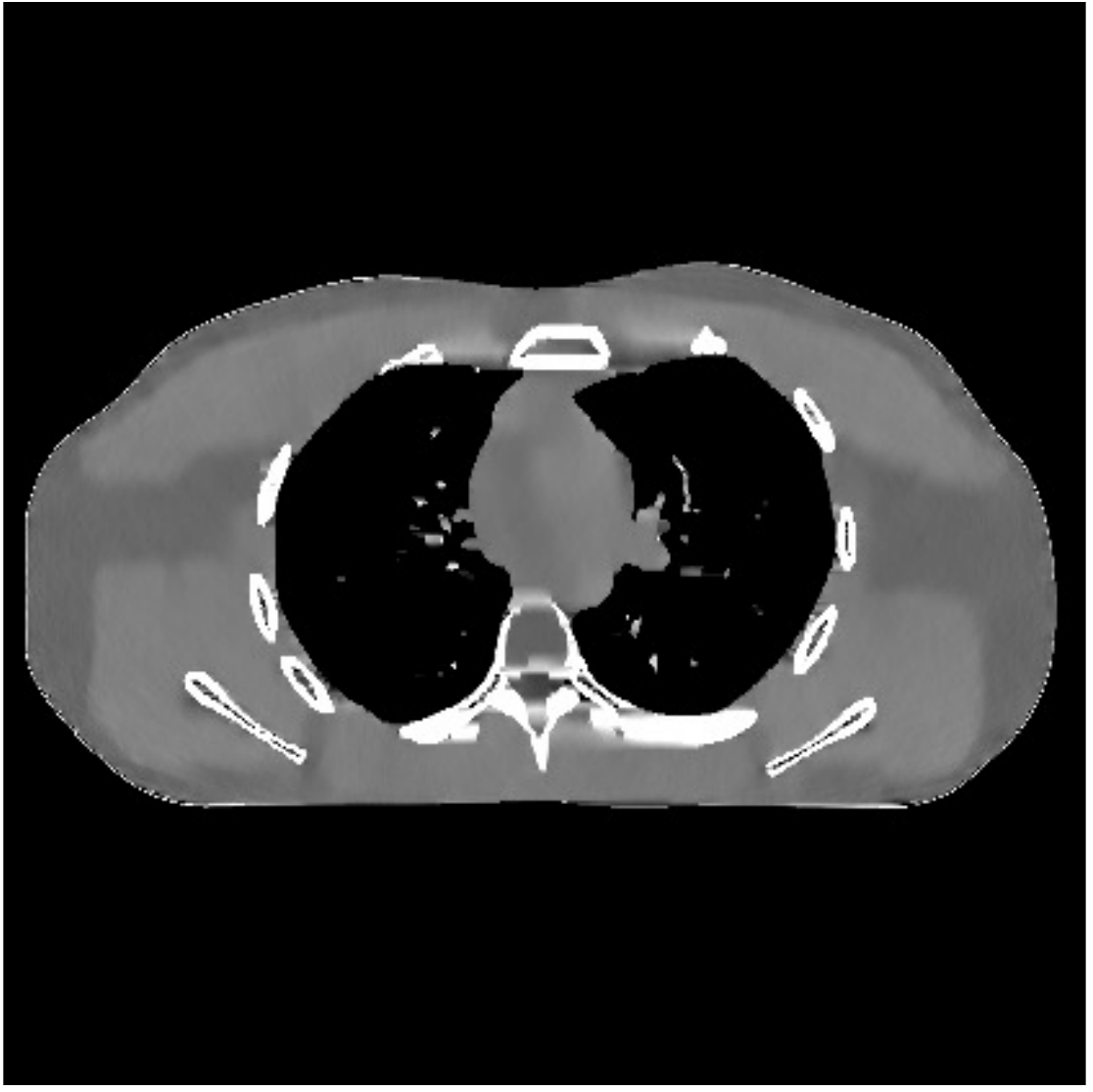}	};
		\spy on (0,0.2) in node [left] at (2,-1.3);
		\node [white] at (0,1.9) {\small $\mathrm{RMSE} = 43.9$,};
		\node [white] at (-.05,1.5) {\small $\mathrm{PSNR} = 31.9$};
		\end{tikzpicture} & \hspace{-0.3in}
		\begin{tikzpicture}
		[spy using outlines={rectangle,green,magnification=2,size=13mm, connect spies}]
		\node {	\includegraphics[width=0.24\textwidth]{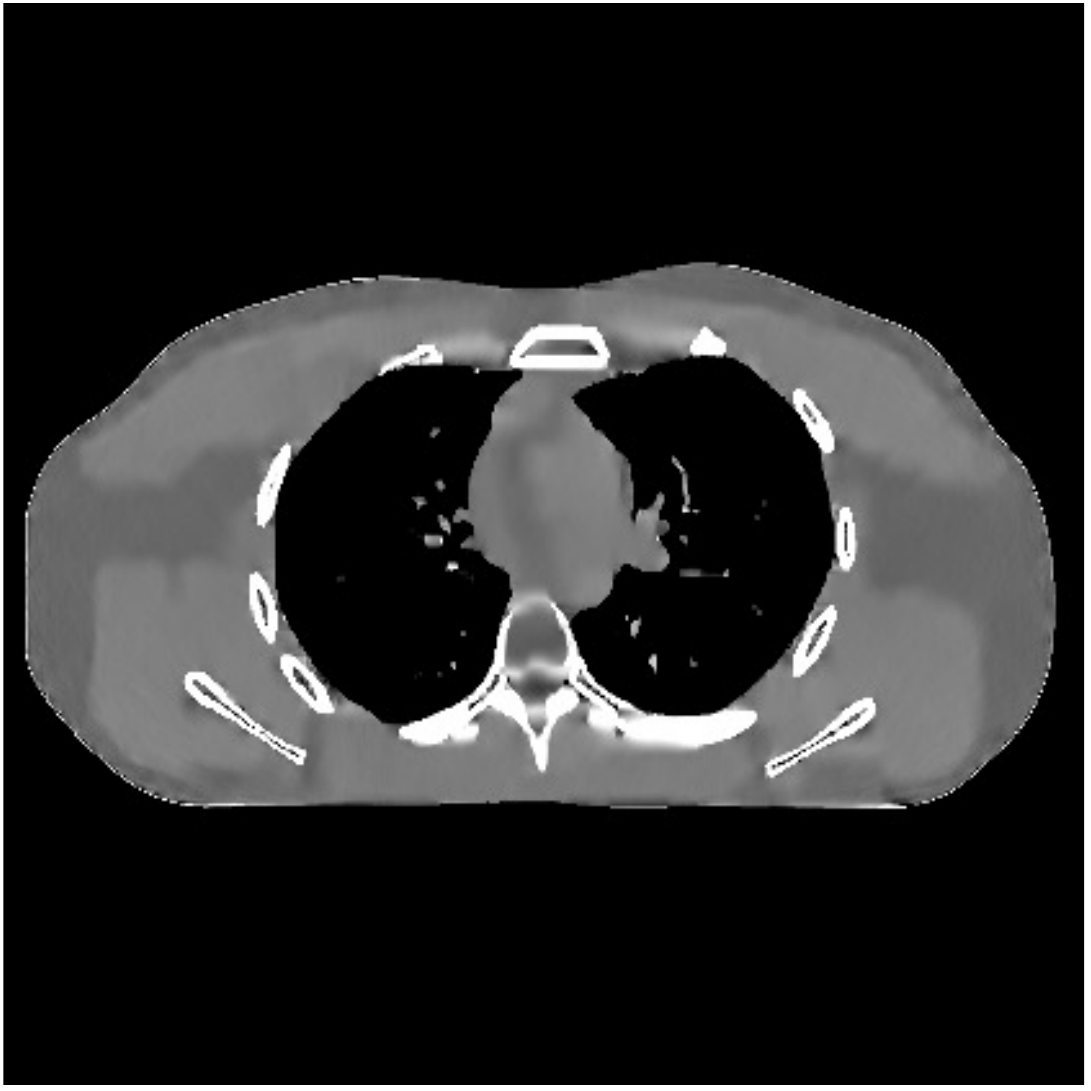}		};
		\spy on (0,0.2) in node [left] at (2,-1.3);	
		\node [white] at (0,1.9) {\small \green{$\mathrm{RMSE} = 42.7$,}};
		\node [white] at (-.05,1.5) {\small \green{$\mathrm{PSNR} = 32.0$}};
		\end{tikzpicture}  \\
		
		(a) FBP &  \hspace{-0.3in}  (b) PWLS-EP  &  \hspace{-0.3in}   (c) PWLS-ST   &   \hspace{-0.3in} (d) PWLS-MRST2 \\
		\vspace{-0.2in}
	\end{tabular}
	
	\caption{Comparison of reconstructions of Slice~$1$ for FBP, PWLS-EP, PWLS-ST, and PWLS-MRST2, respectively at incident photon intensities $I_0 = 5000$. The display window is $[800, 1200]$ HU.}
	\label{fig:2d_recon}
	\vspace{-0.15in}
\end{figure*}

Fig.~\ref{fig:2d_recon} shows representative reconstructions for FBP, PWLS-EP, PWLS-ST, and PWLS-MRST2. 
Compared to FBP and PWLS-EP, PWLS-MRST2 significantly improves image quality by reducing noise and preserving structural details. Furthermore, PWLS-MRST2 improves the quality of the central region and image edges compared to PWLS-ST. 

\begin{table}[!t]	
	\centering
	\caption{RMSE in HU (first row) and PSNR in dB (second row) of fan-beam reconstructions with FBP, PWLS-EP, PWLS-ST, and PWLS-MRST2 for two slices and three incident photon intensities.}
	\label{tab:2d}	 	
	\vspace{-0.05in}
	
	\footnotesize{
		\begin{tabular}{c|ccc|ccc}		
			\toprule
			&	&	Slice~$1$  &  &   &    Slice~$2$ \\
			\midrule
			Intensity   &10000   & 5000   & 3000    &10000   & 5000   & 3000    \\
			\midrule
			\multirow{2}{*} {FBP} & 73.7  & 89.0  & 101.0  & 72.5  & 86.1  & 112.2 \\
			\cmidrule{2-7}
			& 27.3  & 25.7  & 23.5   & 27.5  & 26.0  & 23.7  \\
			\midrule
			\multirow{2}{*} {EP}  & 39.4  & 49.7  & 56.9   & 37.1  & 45.5  & 53.5  \\
			\cmidrule{2-7}
			& 32.8  & 30.8  & 29.6   & 33.3  & 31.5  & 30.1  \\
			\midrule
			\multirow{2}{*} {ST}  & 36.5  & 43.9  & 49.4   & 33.7  & 41.5  & 49.0  \\
			\cmidrule{2-7}
			& 33.4  & 31.9  & 30.8   & 34.1  & 32.3  & 30.9  \\
			\midrule
			\multirow{2}{*} {MRST2}  & $\bm{35.7}$  & $\bm{42.7}$  & $\bm{48.9}$   & $\bm{33.0}$  & $\bm{40.8}$ & $\bm{48.6}$  \\
			\cmidrule{2-7}
			& $\bm{33.6}$  & $\bm{32.0}$  & $\bm{30.9}$   & $\bm{34.3}$  & $\bm{32.5}$ & $\bm{31.0}$  \\
			\bottomrule
		\end{tabular}
	}
	\vspace{-0.15in}
\end{table}

\vspace{-0.15in}
\section{Conclusion}
We presented the learning of a two-layer extension of the sparsifying transform model for CT image reconstruction from low-dose measurements. The model is learned from datasets to sparsify the filtering or transform domain residuals in the second layer.
The algorithms for both learning and reconstruction derived for the simple two-layer case are block coordinate descent-type algorithms and involve efficient updates. Our experimental results illustrated the superior performance of a learned two-layer scheme over the single layer adaptive transform scheme. 
The learned approaches significantly outperformed nonadaptive methods. Since \emph{unsupervised model-based} learning of deep models for imaging is a new area, we plan to investigate the learning of more complex models and more layers for CT image reconstruction and other tasks in future work.

\vspace{-0.15in}
\subsubsection*{Acknowledgments}
\vspace{-0.05in}
The authors thank Xikai Yang for his help with part of the experiments.

\vspace{-0.1in}
\bibliographystyle{IEEEbib}
\bibliography{refs}

\end{document}